\documentclass[prl,twocolumn,10pt,aps,longbibliography,superscriptaddress]{revtex4-1}
\usepackage[utf8]{inputenc}
\usepackage{graphicx}
\usepackage{color}
\usepackage{appendix}
\usepackage{amsmath}
\usepackage{amssymb}
\usepackage{subfigure}
\usepackage{epsfig}
\usepackage{ulem}

\newcommand{\rb}{\boldsymbol{r}}

\begin{document}

\title{Particle Conservation in Dynamical Density Functional Theory}

\author{Daniel de las Heras}
\email{delasheras.daniel@gmail.com}
\affiliation{Theoretische Physik II, Physikalisches Institut, 
  Universit{\"a}t Bayreuth, D-95440 Bayreuth, Germany}

\author{Joseph~M.~Brader}
\email{joseph.brader@unifr.ch}
\affiliation{Department of Physics, University of Fribourg, 
  CH-1700 Fribourg, Switzerland}

\author{Andrea Fortini}
\email{andrea.fortini@me.com}
\affiliation{Theoretische Physik II, Physikalisches Institut, 
  Universit{\"a}t Bayreuth, D-95440 Bayreuth, Germany}
\affiliation{Department of Physics, University of Surrey, 
Guildford GU2 7XH, United Kingdom}

\author{Matthias Schmidt}
\email{matthias.schmidt@uni-bayreuth.de}
\affiliation{Theoretische Physik II,
  Physikalisches Institut, Universit{\"a}t Bayreuth, D-95440 Bayreuth,
  Germany}

\date{1 October 2015}

\begin{abstract}
We present the exact adiabatic theory for the dynamics of the
inhomogeneous density distribution of a classical fluid. Erroneous
particle number fluctuations of dynamical density functional theory
are absent, both for canonical and grand canonical initial
conditions. We obtain the canonical free energy functional, which
yields the adiabatic interparticle forces of overdamped Brownian
motion. Using an exact and one of the most advanced approximate 
hard core free energy functionals, we obtain excellent agreement with
simulations. The theory applies to finite systems in and out of
equilibrium. 
\end{abstract}

\maketitle

Classical density functional theory (DFT) is a highly successful
approach for the description of equilibrium phenomena in both
inhomogeneous liquids and solids. Conventionally, the theory is
formulated in the grand canonical ensemble, where besides the system
volume $V$ and the temperature $T$, the chemical potential $\mu$ is
prescribed. The number of particles, $N$, fluctuates
\cite{evans_dft,mermin}.  However, fixing $N$ in a finite system, as
is done in the canonical ensemble, can be a much more appropriate
representation of an experimental situation.  Examples of such systems
include colloidal clusters \cite{manoharan} and fluids confined to
closed cavities~\cite{PhysRevLett.79.2466,PhysRevLett.84.1220}.  The
differences between canonical and grand canonical results can be very
significant, see e.g.\ Ref.~\cite{PRLcanonical}.

In order to extend DFT to canonical systems, several insightful
studies have been carried out, such as the perturbation approach of
Refs.\ \cite{PhysRevLett.79.2466,PhysRevLett.84.1220} and recent work
on system-size dependence \cite{chakraborty2015}.  Only very recently,
an exact decomposition procedure was discovered~\cite{PRLcanonical},
which allows to obtain e.g.\ canonical density profiles from
minimization of a grand canonical functional.  While the variational
principle of DFT has been formulated in the canonical ensemble
\cite{0953-8984-13-25-101,PhysRevE.83.061133}, any explicit access to
the canonical free energy functional is not available at present.

Dynamical density functional theory (DDFT) is an extension of DFT to
time-dependent situations, where the underlying many-body system is
governed by overdamped Brownian motion
\cite{DDFTTarazona,DDFTArcher}. The DDFT equation of motion has a
drift-diffusion structure, in which the gradient of the local chemical
potential drives the one-body density.  The former is obtained as the
functional derivative of the grand canonical free energy functional
with respect to the density.  This represents an adiabatic
approximation, and captures spatially non-local correlation effects.
There are a considerable number of successful applications of DDFT, as
compared to simulations and experimental results, such as
e.g.\ spinodal decomposition~\cite{DDFTArcher}, driven colloids in
polymer solutions~\cite{PhysRevE.68.061407}, ultrasoft particles in
external fields~\cite{0953-8984-15-6-102} and colloidal
sedimentation~\cite{PhysRevLett.98.188304}.  However, the formulation
confuses canonical and grand canonical concepts.

Despite the importance of choosing the correct ensemble, and the fact
that the deviations of theoretical results from simulation data are
often attributed to ensemble differences, we are not aware of any
systematic work that would address this issue.  Clarifying this
situation has become of particular importance, as recently the
``super-adiabatic'' forces, which are the contributions that are not
derivable from any (adiabatic) free energy, were shown to be highly
non-trivial by explicit many-body simulations~\cite{PRLsuperad}.  A
recent variational approach was formulated that allows to obtain the
``missing'' super-adiabatic forces from functional differentiation of
a free power functional \cite{PowerF}.  To construct theories of the
super-adiabatic forces, which are in general both nonlocal in space
and time, it is important to clarify the issue of ensemble difference.

In this special issue contribution we formulate the correct adiabatic dynamics, which
consistently conserves the number of particles during the time
evolution of the one-body density. 
This enables a systematic study of the dynamics of small systems 
and thus opens a path for the theoretical investigation of 
problems such as, e.g. cluster formation 
or dynamics under confinement. Moreover, 
we show that the
internal adiabatic forces are governed by the canonical free energy
functional $F_N$, and give an explicit method for constructing $F_N$.

First we recall some statistical mechanics. In equilibrium the grand
partition function is
\begin{equation}
\Xi(\mu,V,T) = \sum_{N=0}^\infty e^{\beta\mu N}Z_N(V,T),\label{eq:GCpf}
\end{equation}
where $\beta=1/(k_BT)$, with $k_B$ the Boltzmann constant, and $Z_N$
the canonical partition function of a system with $N$ particles 
The thermodynamic grand potential is
\begin{equation}
\Omega_0=-k_BT\ln\Xi\label{eq:GP}.
\end{equation}
Equilibrium grand canonical density profiles, $\rho_\mu(\rb)$, are a
direct result of the DFT minimization for given value of $\mu$, and
are related to the canonical profiles $\rho_N(\rb)$ via
\begin{equation}\label{densitysum}
  \rho_\mu(\rb) = \sum_N p_N(\mu)\rho_N(\rb),
\end{equation}
where the probability $p_N(\mu)$ of finding $N$ particles at given
chemical potential $\mu$ is
\begin{equation}\label{eq:pNdefinition}
  p_N(\mu)=\exp(\beta\mu N)\frac {Z_N}{\Xi(\mu)}.
\end{equation}
The decomposition method of Ref.\ \cite{PRLcanonical} amounts to
choosing an appropriate set of values of the chemical potential,
$\{\mu_1,\ldots\mu_{N_{\rm max}}\}\equiv \{\mu_n\}$ and regarding
$p_N(\mu_n)$ as the $(N,n)$ element of an $N_\text{max}\times
N_\text{max}$ matrix, ${\bf P}$. Here $N_{\rm max}$ is an upper cutoff
in \eqref{densitysum} and the trivial case $N=0$ has been removed
\cite{PRLcanonical}. The matrix ${\bf P}$ can be constructed from DFT
results for $\Omega_0(\mu_n)$, obtained for all $\{\mu_n\}$, and
solving the resulting system of linear equations \eqref{eq:GP} and
\eqref{eq:pNdefinition} for the set $Z_N$ and hence $p_N(\mu_n)$.  The
inverse matrix ${\bf P}^{-1}$, with elements ${\bf P}^{-1}_{\!Nn}$,
can then be used to decompose any grand ensemble average into the
underlying canonical contributions.  For example, the canonical
density profiles are given by
\begin{equation}
  \rho_N(\rb)=  \sum_{n}{\bf P}^{-1}_{\!Nn}  \rho_{\mu_n}(\rb).
  \label{invertrho}
\end{equation}

Access to the canonical free energy functional $F_N[\rho]$ is not
presently available.  In order to provide this, let $\rho(\rb)$ be an
arbitrary trial canonical density profile, with fixed number of
particles, $\int d\rb\rho(\rb)=N$.  We turn~$\rho(\rb)$ into the
target for an inversion procedure to find the corresponding external
potential $V(\rb)$, that generates $\rho(\rb)$ in (canonical)
equilibrium. Then by subtracting the external contribution to the
canonical free energy, the value of the canonical intrinsic free
energy functional, $F_N$, evaluated at $\rho(\rb)$, can be obtained
via
\begin{equation}
  F_N[\rho]=
  -k_{\rm B}T\ln Z_N
  -\int d\rb \rho(\rb)V(\rb).
\label{cDFT}
\end{equation}

In order to find $V(\rb)$, we start with the grand canonical
Euler-Lagrange equation:
\begin{equation}
  \beta V(\rb) =  c^{(1)}_\mu(\rb) +\beta\mu -\ln\rho_\mu(\rb),\label{Euler}
\end{equation}
where $c^{(1)}_\mu(\rb)$ is the one-body direct correlation function
for density profile $\rho_\mu(\rb)$ and we have set the irrelevant
thermal wavelength to unity.  We have developed the following
efficient iteration scheme. We start with an initial guess
$V^{(0)}(\rb)$ and define the $i$th iteration step via
\begin{equation}
  \label{iteration}
  \beta V^{(i)}(\rb) = 
  \beta V^{(i-1)}(\rb)
  -\ln\rho(\rb)
  +\ln\sum_n{\bf P}^{-1}_{\!Nn}
  \rho_{\mu_n}(\rb),
\end{equation}
which can be derived from inserting Eq.~(\ref{densitysum}) into
(\ref{Euler}) and then inverting with (\ref{invertrho}).  The terms in
the sum in Eq.~\eqref{iteration} are re-calculated at each step, using
the decomposition procedure described above.

\begin{figure}[ht]
\includegraphics[width=0.9\columnwidth]{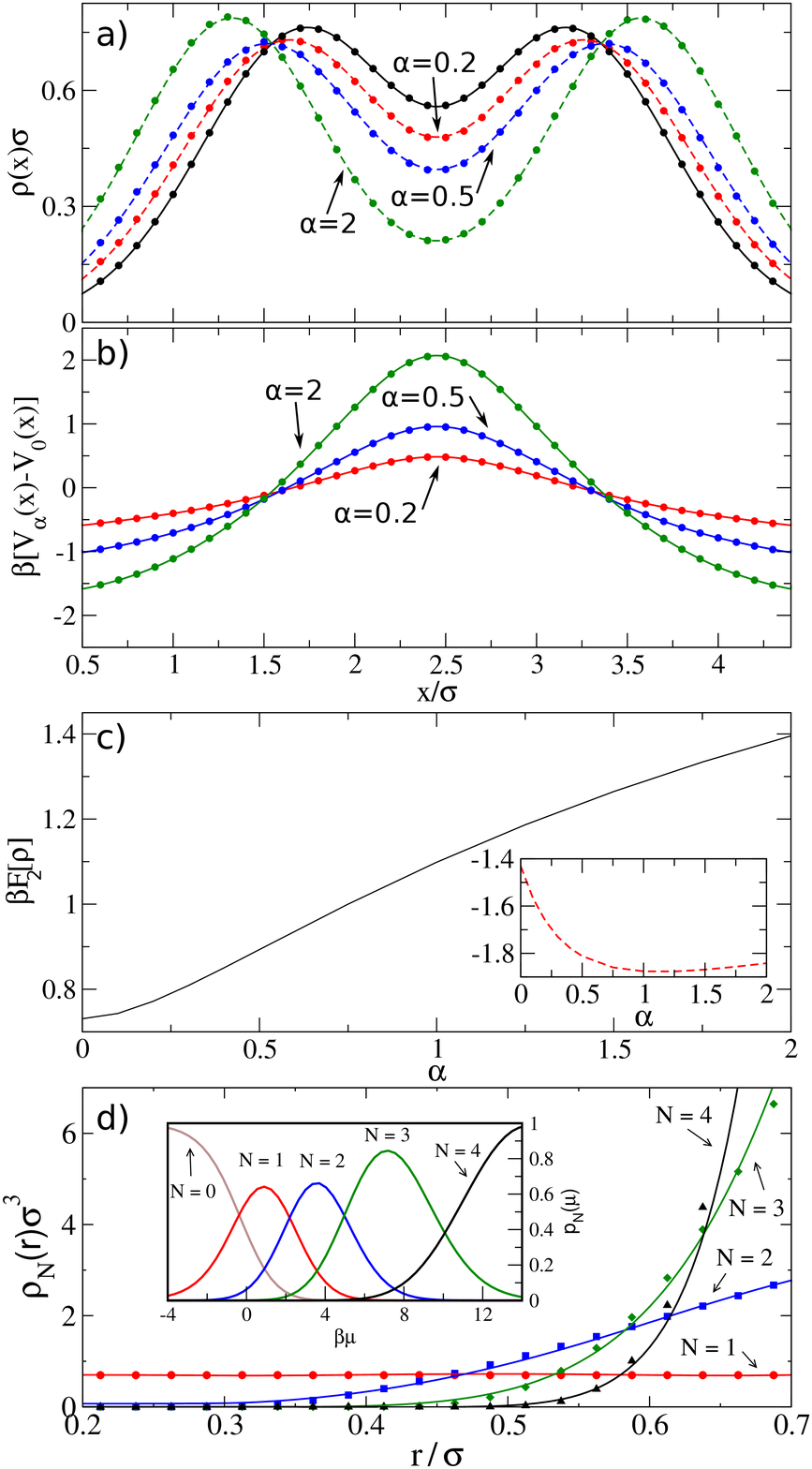}
\caption{\label{fig1} (a) Equilibrium density profile $\rho_{N=2}(x)$
  (solid line) of a system of $N=2$ hard rods confined in a slit pore
  and in presence of a parabolic external potential. The dashed lines
  are trial profiles obtained via
  $\rho(x)=\rho_{N=2}(x)A[1+\alpha(x-h/2)^2]$ for different
  perturbation strengths $\alpha$, as indicated. (b) Difference of
  scaled external potentials $\beta(V_\alpha(x)-V_0(x))$ for different
  values of $\alpha$. Symbols in (a) and (b) correspond to Monte Carlo
  simulation, using the inversion procedure of Ref.~\cite{PRLsuperad}
  to obtain $V_\alpha(x)$.  (c) Value of the canonical functional
  $\beta F_{N=2}$ and the intrinsic grand canonical free energy
  functional ~\cite{percus} (inset) evaluated at $\rho$ as a function
  of the perturbation strength $\alpha$. (d) Canonical density
  profiles of hard spheres confined in a spherical cavity of radius
  $r/\sigma=1.2$. The solid lines are obtained via decomposition of
  the grand canonical functional White Bear mk.~II ~\cite{WB2}.
  Symbols represent Monte Carlo simulation data. The inset shows the
  probabilities $p_N$ as a function of $\mu$ for $N=1,2,3,4$, as
  indicated.  }
\end{figure}

We first apply the method to a system of one-dimensional hard
particles, for which the exact grand canonical (Helmholtz) intrinsic free energy
functional $F[\rho_\mu]$ is known \cite{percus}.  In order to provide
a severe test of the canonical functional approach, we consider $N=2$
particles of length $\sigma$ confined between two identical hard walls
separated by a distance $h=4.9\sigma$ along the $x$-axis.  In addition
we apply a parabolic external potential
$V_0(x)=(x-h/2)^2k_BT/\sigma^{2}$.  First we find the equilibrium
canonical profile $\rho_{N=2}(x)$ using Eq.~\eqref{invertrho}.  Next,
we generate trial density profiles $\rho_\alpha(x)$ via a
multiplicative perturbation:
$\rho_\alpha(x)=A[1+\alpha(x-h/2)^2]\rho_{N=2}(x)$, where $A$ is a
constant that normalizes the profile such that it contains two
particles, and $\alpha$ determines the strength of the perturbation,
see Fig.~\ref{fig1}a.  The corresponding external potential
$V_{\alpha}(\rb)$ is then obtained by the iterative
method~\eqref{iteration}. In Fig.~\ref{fig1}b we show results for
$V_\alpha(x)-V_0(x)$ for a range of values of $\alpha$. The value of
the canonical free energy functional, $F_N[\rho_\alpha]$, follows from
Eq.~\eqref{cDFT}; the results are plotted in Fig.~\ref{fig1}c. As
expected, the canonical free energy increases with the perturbation
strength $\alpha$, and it is completely different from the intrinsic Helmholtz
grand canonical free energy~\cite{percus},
see the inset of Fig.~\ref{fig1}c for $F[\rho_\alpha]$. Here
$F[\rho_\alpha]$ consists of the ideal gas functional and Percus' excess
free-energy functional evaluated at $\rho_\alpha$. 

In order to demonstrate the applicability of the method to more
realistic systems, we consider a three-dimensional case of hard
spheres confined in a hard spherical cavity. We employ one of the most
advanced free energy functionals presently available, namely the
tensorial White Bear II fundamental measure functional \cite{WB2}. The
agreement of the canonical density profiles, as compared to Monte
Carlo simulation data, is remarkable, see Fig.~\ref{fig1}d. The inset
of Fig.~\ref{fig1}d shows the probabilities $p_N$ as a function of
$\mu$.

The canonical equilibrium state serves as an initial condition for the
time evolution. To describe the many-body dynamics, we employ the
$N$-particle Smoluchowski equation \cite{DDFTArcher}, which locally
conserves the particles throughout the time evolution (no exchange
with any particle bath). An exact equation of motion for the
time-dependent density profile $\rho_N(\rb,t)$ is obtained by
integrating over $N\!-\!1$ degrees of freedom,
\begin{align}\label{exacteom}
  \frac{\partial \rho_N(\rb,t)}{\partial t}
  &= D_0\nabla\cdot \Big[\nabla\rho_N(\rb,t)
    - \beta {\bf f}_N(\rb,t)\\
    &\qquad\qquad - \beta\rho_N(\rb,t)
    ({\bf X}(\rb,t)-\nabla V_{\rm ext}(\rb,t))
    \Big], \notag
\end{align}
where $D_0$ is the bare diffusion coefficient, $V_{\rm ext}(\rb,t)$ is
a time-dependent external potential, ${\bf X}(\rb,t)$ is a
non-conservative force field, and ${\bf f}_N(\rb,t)$ is the internal
force density due to the interparticle interactions.  The latter is
given exactly by
\begin{align}\label{force_density}
  {\bf f}_N(\rb,t) = 
  -\int d\rb'\, \rho_N^{(2)}(\rb,\rb',t)\nabla u(|\rb-\rb'|)
\end{align}
where $\rho_N^{(2)}(\rb,\rb',t)$ is the exact nonequilibrium pair
density for $N$ particles, and $u(r)$ is the interparticle pair
potential. Schmidt and Brader \cite{PowerF} have shown that the
internal force density can be systematically split into an adiabatic
and a superadiabatic contribution,
\begin{align}
  {\bf f}_N(\rb,t) &= {\bf f}_N^{\rm ad}(\rb,[\rho_N])
  + {\bf f}_N^{\rm sup}(\rb,t),
\end{align}
where the adiabatic force density is an instantaneous functional of
the one-body density distribution and ${\bf f}_N^{\rm sup}(\rb,t)$
contains memory effects, which are neglected in DDFT.  The adiabatic
approximation corresponds to setting ${\bf f}_N^{\rm sup}(\rb,t)=0$; a
fundamental assumption of DDFT, which we retain in the present work.
In contrast to DDFT, however, we will treat ${\bf f}_N^{\rm ad}(\rb)$
exactly.

The instantaneous nonequilibrium density $\rho_N(\rb,t)$ allows to
define at each time $t$ an adiabatic reference state as an equilibrium canonical
ensemble of $N$ particles with one-body density distribution
\begin{align}
  \label{EQadiabaticCanonicalDensity}
  \rho_N^{\rm ad}(\rb)=\rho_N(\rb,t).
\end{align}
Here the left hand side (as well as all subsequent adiabatic quantities)
is in general different at each time. We suppress this time dependence in the notation
in order to highlight the static nature of the adiabatic state. 
The canonical inversion procedure \eqref{iteration} then determines
the corresponding external (``adiabatic'') potential, $V_{\rm
  ad}(\rb)$, which, together with $u(r)$, specifies the adiabatic
system completely. Note that $V_{\rm ad}(\rb)$ is in general unrelated
to $V_{\rm ext}(\rb,t)$ (as occurring in the equation of motion
\eqref{exacteom}).  The corresponding canonical two-body density
distribution $\rho_N^{(2){\rm ad}}(\rb,\rb')$ and the internal force
density in the adiabatic system are related by
\begin{align}
  {\bf f}_N^{\rm ad}(\rb) &= 
  -\int d\rb' \rho_N^{(2)\rm ad}(\rb,\rb')\nabla u(|\rb-\rb'|).
\end{align}
In the following we demonstrate how ${\bf f}_N^{\rm ad}(\rb)$ can be
explicitly calculated. This specifies the adiabatic one-body dynamics
completely.  We present three different alternatives, all of which
yield the same result.

i)~As the adiabatic system is in equilibrium, the net force vanishes.
Hence the internal forces equal the negative external and entropic
forces, and
\begin{align}
  {\bf f}_N^{\rm ad}(\rb) &= \rho_N^{\rm ad}(\rb)\nabla 
  [V_{\rm ad}(\rb) + k_BT \ln \rho_N^{\rm ad}(\rb)].
\end{align}
Here all quantities on the right hand side are known: the adiabatic
density $\rho_N^{\rm ad}(\rb)$ via
\eqref{EQadiabaticCanonicalDensity}, and $V_{\rm ad}(\rb)$ has already
been obtained from the canonical inversion procedure.

ii) Functional differentiation of the canonical excess (over ideal gas)
free energy functional $F_N^{\rm exc}[\rho]$ yields
\begin{align}
  {\bf f}_N^{\rm ad}(\rb) &= - \rho_N^{\rm ad}(\rb)\nabla
  \left.  \frac{\delta F^{\rm exc}_N[\rho]}{\delta \rho(\rb)} 
  \right|_{\rho(\rb)=\rho_N^{\rm ad}(\rb)}.
\end{align}
In practice this procedure requires performing the functional
derivative numerically.

\begin{figure}
\includegraphics[width=0.9\columnwidth]{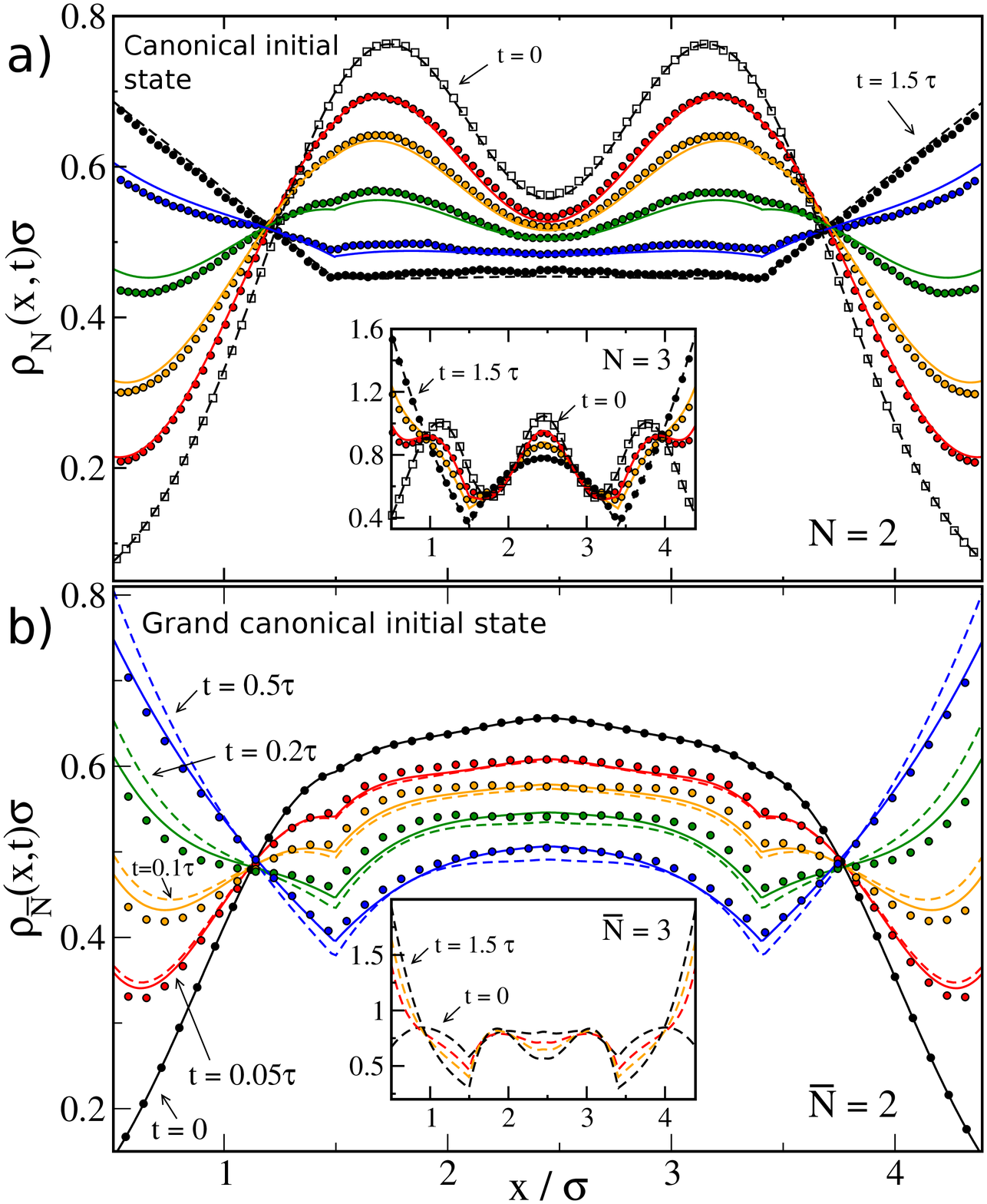}
\caption{\label{fig2} {(a) Time evolution of canonical density
    profiles for a system of $N=2$ and $N=3$ (inset) particles in one
    dimension confined to a slit of width $h=4.9\sigma$.  For $t<0$
    the external potential consists of a harmonic trap $V_{\rm
      ext}(x)=(x-h/2)^2k_BT/\sigma^2$ and hard walls at $x=0$ and
    $x/\sigma=4.9$ (such that the density is cut at $x/\sigma=0.5$ and
    $4.4$).  At $t=0$ the harmonic trap is switched off and the
    density relaxes.  The density at $t=0$ and $t=1.5\tau$ are given
    by the dashed lines, as indicated; the time scale is
    $\tau=\sigma^2/D_0$. At $t=1.5\tau$ the system has practically
    relaxed to the final equilibrium state.  Intermediate
    nonequilibrium profiles are shown at times $t/\tau=0.05, 0.10,
    0.20, 0.40$ (for $N=2$) and $t/\tau=0.05, 0.10, 0.15$ (for $N=3$)
    and are given by the full lines.  Symbols indicate the results of
    Brownian dynamics simulations.  (b) Same as panel (a), but for an
    initial grand canonical profile with average number of particles
    is $\overline N=2$, according to DDFT (dashed lines), the particle
    conserving theory (solid lines) and Brownian dynamics simulation
    (symbols). Solid lines and symbols have been obtained by
    recomposition of canonical states according to
    Eq.~\eqref{densitySuperposition}. The initial state, for $t=0$, is
    the same in all cases.  At $t=0.5\tau$ the system has almost
    relaxed to its final state. The inset in (b) shows the time
    evolution according to DDFT of a grand canonical profile with
    $\overline N=3$.}}
\end{figure}

iii) From decomposition of the force in an adiabatic grand canonical
state one obtains
\begin{equation}
  {\bf f}_N^{\rm ad}(\rb)=k_BT
  \sum_n{\bf P}^{-1}_{\!Nn}
  \rho_{\mu_n}(\rb)\nabla c^{(1)}_{\mu_n}(\rb),
  \label{invertf}
\end{equation}
where $\rho_{\mu_n}^{\rm ad}(\rb)$ is a set of grand canonical density
profiles in the adiabatic potential $V_{\rm ad}(\rb)$, and
$c^{(1)}_{\mu_n}(\rb)$ are the corresponding one-body direct
correlation functions. Eq.~\eqref{invertf} can be derived from
the exact grand canonical sum rule
\begin{align}
  \label{sumrule}
  k_BT\rho_\mu(\rb) \nabla c^{(1)}_\mu(\rb)
  &= -\int d\rb'\rho_\mu^{(2)\rm ad}(\rb,\rb')\nabla u(|\rb-\rb'|),
\end{align}
and decomposing the grand canonical two-body density $\rho_\mu^{(2)\rm
  ad}(\rb,\rb')$ in the adiabatic system. We have explicitly 
verified that the three methods yield the same results within numerical
accuracy.

We are now in a position to integrate Eq.~\eqref{exacteom} in time
using purely canonical forces to drive the dynamics.  The adiabatic
potential is re-calculated at each time step.  In Fig.~\ref{fig2}a we
show results from the particle conserving theory for the relaxation of
$N=2$ and $N=3$ (inset) hard rods, following the switching-off of a
harmonic potential.  The rods remain confined between two hard walls
for all times and each system relaxes to its final (canonical)
equilibrium state. The theoretical results are in very good agreement
with our Brownian Dynamics simulation results 
(simulation details can be found in \cite{PRLsuperad}). 

The theoretical time
evolution is slightly ahead of the simulation data.  This is
consistent with the direction of the super-adiabatic forces, which we
have obtained by simulations, following the method of
Ref.\ \cite{PRLsuperad}; these results will be presented elsewhere.
For the dense state $N=4$ we also find very good agreement of
theoretical results and simulation data (not shown). Any systematic
deviations of theoretical results from the simulation data are
entirely due to the omission of super-adiabatic forces in the theory,
and not due to ensemble differences.  For $N=1$ the theory is exact,
as the super-adiabatic forces vanish.

It is now straightforward to generalize to grand canonical initial
conditions. Let the system at the initial time $t=0$ be specified by a
grand canonical density distribution $\rho_\mu^{(0)}(\rb)$ with
average number of particles $\overline N = \sum_N Np_N^{(0)}(\mu)$.
This state can be viewed as being composed of a set of underlying
canonical density profiles $\rho_N^{(0)}(\rb)$ with statistical
weights $p_N^{(0)}(\mu)$.  Each of these canonical states evolves in
time under particle conserving dynamics. Hence the entire grand
canonical initial state evolves as a superposition of the trajectories
$\rho_N(\rb,t>0)$. The statistical weights, however, are those of the
initial grand canonical state, $p_N^{(0)}(\mu)$, as the system is
decoupled from any particle bath for $t>0$ (there is no source
  term in \eqref{exacteom}). Hence the one-body density of this system is
given by
\begin{align}
  \rho_{\overline N}(\rb,t) = \sum_N p_N^{(0)}(\mu) \rho_N(\rb,t).
  \label{densitySuperposition}
\end{align}

Fig.~\ref{fig2}b shows corresponding results for $\overline N=2$ and
$\overline N=3$ (inset). We find again very good agreement between the
theory and BD simulation data.  The theoretical time evolution is
slightly ahead of the BD data, which is entirely due to having
  neglected super-adiabatic forces in the theory. 
The theory captures the correct long-time limit. 
The time evolution of these
initially grand canonical states differs very significantly from that
of the corresponding canonical initial states, shown in
Fig.~\ref{fig2}a. This striking discrepancy occurs despite the fact
that $\overline N=N$,
which highlights the importance of correct choice of ensemble in
finite systems.

We next compare our approach to DDFT. As demonstrated by Archer and
Evans \cite{DDFTArcher}, DDFT amounts to employing the equilibrium sum
rule \eqref{sumrule} for expressing the interaction force in terms of
the one-body direct correlation function in the grand
ensemble. However, instead of using the correct
  relation~\eqref{EQadiabaticCanonicalDensity}, 
DDFT amounts to constructing a grand canonical adiabatic
  state, with density distribution

\begin{align}
  \rho_\mu^{\rm ad}(\rb) &= \rho_N(\rb,t).
\end{align} 
Via the Euler-Lagrange equation \eqref{Euler}, a corresponding
external potential exists, that generates $\rho_\mu^{\rm ad}(\rb)$ in
the grand ensemble.
The grand canonical adiabatic system under the influence of this
external potential possess a two-body density,
$\rho_\mu^{(2)}(\rb,\rb')$, for which using the sum rule \eqref{sumrule} yields 
 the associated force density
\begin{align}
{\bf f}_{\rm DDFT}(\rb)=
  -k_BT \rho_\mu^{\rm ad}(\rb) \nabla c^{(1)}_\mu(\rb),
\end{align}
which differs from the exact expression \eqref{invertf}. 
In the example of Fig.~\ref{fig2}b, although DDFT deviates more
strongly from the simulation data than the present theory, it
nevertheless provides a reasonable description of the dynamics of an
initial grand canonical state.

Although we have presented results for very simple test cases, the
particle conserving dynamical theory is applicable to 
any system for which a grand canonical density functional is available.
Studies of complex phenomena, such as the dynamics of colloidal
cluster formation, or transport through ion channels are thus within
reach.  As exemplified by the comparison of Fig.~\ref{fig2}a and 2b,
the time evolution of a system containing only a few particles is very
sensitive to the choice of ensemble. In systems with a reduced number
of particles the use of a canonical DFT and particle conserving
dynamics is indispensable in order to compare with experiments or
simulations performed at fixed particle number. Canonical and 
grand canonical ensembles are equivalent in the thermodynamic limit, and
the time evolution in DFT is just a temporal sequence of equilibrium states. 
Hence, one might expect our particle conserving theory and  (standard) DDFT to be 
equivalent in systems with a large number of particles. However, local
fluctuations typically involve only a reduced number of particles. Therefore, the dynamics
of localized phenomena might depend on the ensemble, even in the 
thermodynamic limit. This is an open problem to be addressed in future work.

DdlH and JMB contributed equally to this work.

\end{document}